\documentclass[runningheads]{llncs}
\usepackage{booktabs}
\usepackage{multirow}
\usepackage{graphicx}
\usepackage{url}
\graphicspath{{fig/}}
\usepackage{amsmath}
\usepackage{hyperref}
\usepackage[dvipsnames]{xcolor}

\DeclareMathOperator{\sgn}{sgn}

\begin{document}
%
\title{MP3 Compression To Diminish Adversarial Noise in End-to-End Speech Recognition}
%
\titlerunning{MP3 Compression To Diminish Adversarial Noise}
%
\author{Iustina Andronic\inst{1} \and
Ludwig K{\"u}rzinger\inst{2} \and
Edgar Ricardo Chavez Rosas\inst{2} \and
Gerhard Rigoll\inst{2} \and
Bernhard U. Seeber\inst{1}
} 

\authorrunning{I. Andronic et al.}
%
\institute{Audio Information Processing,\and
Chair of Human-Machine Communication,\\ Department of Electrical and Computer Engineering,\\ Technical University of Munich, Germany 
\email{\{iustina.andronic,ludwig.kuerzinger\}@tum.de}
}
\maketitle              
\begin{abstract}

Audio Adversarial Examples (AAE) represent specially created inputs meant to trick Automatic Speech Recognition (ASR) systems into misclassification.
The present work proposes MP3 compression as a means to decrease the impact of Adversarial Noise (AN) in audio samples transcribed by ASR systems.
To this end, we generated
AAEs with the Fast Gradient Sign Method for an end-to-end, hybrid CTC-attention ASR system.
Our method is then validated by two objective indicators:
(1) Character Error Rates (CER) that measure the speech decoding performance of four ASR models trained on uncompressed, as well as MP3-compressed data sets and
(2) Signal-to-Noise Ratio (SNR) estimated for both uncompressed and MP3-compressed AAEs that are reconstructed in the time domain by feature inversion.

We found that MP3 compression applied to AAEs indeed reduces the CER when compared to uncompressed AAEs.
Moreover, feature-in\-verted (reconstructed) AAEs had significantly higher SNRs after MP3 compression, indicating that AN was reduced.
In contrast to AN, MP3 compression applied to utterances augmented with regular noise resulted in more transcription errors, giving further evidence that MP3 encoding is effective in diminishing only AN.

\end{abstract}

\section{Introduction}
In our increasingly digitized world,  Automatic Speech Recognition (ASR) has become a natural and convenient means of communication with many daily-use gadgets.
Recent advances in ASR push toward end-to-end systems that directly infer text from audio features, with sentence transcription being done in a single stage, without intermediate representations ~\cite{Watanabe2017,hannun2014deep}.
Those systems do not require handcrafted linguistic information, but learn to extract this information by themselves, as they are trained in an end-to-end manner.
At present, there is a trend towards Deeper Neural Networks (DNN), however these ASR models are also more complex and prone to security threats:
Audio Adversarial Examples (AAEs) are audio inputs which carry along a hidden message induced by adding Adversarial Noise (AN) to a regular speech input.
AN is optimized so that it will mislead the ASR system into misclassification (i.e., recognizing the hidden message), while the AN itself is supposed to remain inconspicuous to humans~\cite{carlini2016hidden,iter2017generating,Alzantot2018,Schonherr2019a}.
Yet from the perspective of ASR research, the system should classify the sentence as close as it is understood by humans.
For this reason, we proceed to investigate MP3 compression as a means to diminish the detrimental effects of AN to ASR performance.
\textbf{Our contributions are three-fold:}
\begin{itemize} 
	\item We create AAEs in the audio domain via a feature inversion procedure;
	\item We evaluate MP3's effectiveness to diminish AN with four end-to-end ASR models trained on different levels of MP3 compression that decode AAEs in uncompressed and MP3 formats derived from the VoxForge corpus;
	\item Conversely, we assess the effects of MP3 compression when applied to inputs augmented with regular, non-adversarial noise.
\end{itemize}

\section{Related Work}

Adversarial Examples (AEs) were first created for image classification tasks~\cite{Goodfellow2015}.
Regular gradient-based training of DNNs calculates the gradient of a chosen loss function w.r.t. the networks' parameters, aiming for their step-wise gradual improvement. 
By contrast, the Fast Gradient Sign Method (FGSM) \cite{Goodfellow2015} creates AN based on the gradient w.r.t. the \emph{input data} in order to optimize towards misclassification.
FGSM was already applied in the context of end-to-end ASR to DeepSpeech~\cite{hannun2014deep,Carlini2018-speech2text}, a CTC-based speech recognition system, as well as for the attention-based system called Listen-Attend-Spell (LAS)~\cite{chan2016listen,Sun2019}. 

For generating AAEs, most of the previous works have set about to pursue either one of the following goals, sometimes succeeding in both:
(1) that their AAEs work in a physical environment and (2) that they are less perceptible to humans.
Carlini et al. \cite{carlini2016hidden} were the first to introduce the so-called \textit{hidden voice commands}, demonstrating that targeted attacks operating over-the-air against archetypal ASR systems (i.e., solely based on Hidden Markov Models - HMM), are feasible.
In contrast to previous works that targeted short adversarial phrases, \cite{Carlini2018-speech2text} constructed adversarial perturbations for designated longer sentences.
Their novel attack was achieved with a gradient-descent minimization based on the CTC loss function, which is optimized for time sequences.
Moreover, \cite{Schonherr2019a} were the first to develop imperceptible AAEs for a conventional hybrid DNN-HMM ASR system by leveraging human psychoacoustics, i.e., manipulating the acoustic signal below the thresholds of human perception.
In a follow-up publication \cite{Schonherr2019b}, their \textit{psychoacoustic hiding} method was enhanced to produce \textit{generic} AAEs that remained \textit{robust} in simulated over-the-air attacks.

Protecting ASR systems against AN that is embedded in AEs has also been primarily investigated in the image domain, which in turn inspired research in the audio domain.
Two major defense strategies are considered from a security-related perspective~\cite{Hu2019}, namely \textit{proactive} and \textit{reactive} approaches.
The former aims to enhance the robustness during the training procedure of the ASR models themselves, e.g., by adversarial training \cite{Sun2019} or network distillation \cite{papernot2016distillation}.
Reactive approaches instead aim to detect if an input is adversarial after the DNNs are trained, by means of e.g., input transformations such as compression, cropping, resizing (for images), meant to at least partially discard the AN and thus recover the genuine transcription.
Regarding AAEs, primitive signal processing operations such as local smoothing, down-sampling and quantization were applied to input audio so as to disrupt the adversarial perturbations ~\cite{yang2018mitigating}. 
The effectiveness of that pre-processing defense was demonstrated especially for shorter-length AAEs.
Rajaratnam et al. \cite{rajaratnamspeech} likewise explored audio pre-processing methods such as band-pass filtering and compression (MP3 and AAC), while also venturing to more complex speech coding algorithms (Speex, Opus) so as to mitigate the AAE attacks.
Their experiments, which however targeted a much simpler keyword-spotting system with a limited dictionary, indicated that an {ensemble strategy} (made of both speech coding and other form of pre-processing, e.g., compression) is more effective against AAEs. 

\section{Experimental Set-up}
MP3 is an audio compression algorithm that employs a lossy, perceptual audio coding scheme based on a psychoacoustic model
in order to discard audio information below the hearing threshold and thus diminish the file size \cite{Brandenburg1999}.
Sch{\"o}nherr et al. \cite{Schonherr2019a} recently hypothesized that MP3 can make for a robust countermeasure to AAE attacks, as it might remove exactly those inaudible ranges in the audio where the AN lies.
However, to date there is no published experimental work to prove the effectiveness of MP3 compression in mitigating AN targeted at a hybrid, end-to-end ASR system.
Consequently, given an audio utterance that should transcribe to the original, non-adversarial phrase, we formulate our research question as follows:
\emph{To what extent can MP3 aid in removing the AN and thus recover the benign character of the input?} 
Hence, we aim to analyze the AN reduction with two objective indicators: Character Error Rates (CER) and Signal-to-Noise Ratio.

\begin{figure}
\centering
\includegraphics[width=0.95\textwidth]{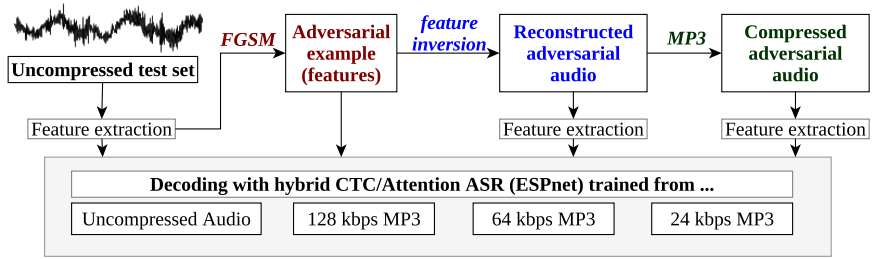}
\caption{General experimental workflow}
\label{fig:workflow}
\end{figure}

\paragraph{\textbf{Pipeline from original audio data to MP3-compressed AAEs.}}
A four-stage pipline that transforms the original test data to MP3 compressed AAEs was implemented;
transformation includes the FGSM method and feature inversion, as depicted in Fig.~\ref{fig:workflow}.
To also consider the effects of MP3 compression on the ASR performance, i.e., whether the neural network adapts to MP3 compression,
experiments were validated on four ASR models \textit{trained} on data with four different levels of MP3 compression (uncompressed, 128 kbps, 64 kbps and 24 kbps MP3).
Format-matched AAEs were then decoded by each of the four models.

All experiments are based on the English share of the open-source VoxForge speech corpus.
It consists of $\approx$130.1 hours of utterances in various English accents that were recorded and released in uncompressed format (\textit{.wav)},
allowing for further compression and thus, for the exploration of our research question.
The Lame MP3 encoder\footnote{https://lame.sourceforge.io/about.php} was used as command line tool for batch MP3 compression.
Log-mel filterbank (fbank) features were extracted
from every utterance of the speech corpus and then fed to the input of the ASR models in both training and testing stages.
The speech recognition experiments are performed with the hybrid CTC-attention ASR system called ESPnet~\cite{Watanabe2017,Watanabe2018}, which combines the two main techniques for end-to-end speech recognition.
First, Connectionist Temporal Classification (CTC~\cite{GravesEtAl06}) carries the concept of hidden Markov states over to end-to-end DNNs as training loss for sequence-classification networks.
DNNs trained with CTC loss classify token probabilities for each time frame.
Second, attention-based encoder-decoder architectures~\cite{chan2016listen} are trained as auto-regressive, sequence-generative models that directly generate the sentence transcription from the entire input utterance.
The multi-objective learning framework unifies the Attention-loss and CTC-loss with a linear interpolation weight (the parameter $\alpha$ from Eq.~\ref{eq:mtl_loss}), which is set to 0.3 in our experiments, allowing attention to dominate over CTC.
Different from the original LAS ASR system \cite{chan2016listen}, we use a {location-aware} attention mechanism that additionally uses attention weights from the previous sequence step;
a more detailed description is to be found in \cite{kurzinger2019exploring}.
\begin{equation}
   Loss_{hybrid} = \alpha \log p_{CTC}(Y|X) + (1-\alpha) \log p_{Att.}(Y|X), \quad \alpha \in [0,1]  \label{eq:mtl_loss}
\end{equation}
Full descriptions of the system's architecture can be consulted in~\cite{Watanabe2017,Watanabe2018,kurzinger2019exploring}.
We use the default ESPnet configuration\footnote{
ESPnet commit \texttt{81a383a9}.
The full parameters' configuration is also listed in Table 4.2 of the Master's Thesis underlying this publication~\cite{Andronic2020}.
} for the VoxForge dataset~\cite{voxforge_misc}.

\begin{figure}[!tb]
  \centering
\includegraphics[width=1\linewidth]{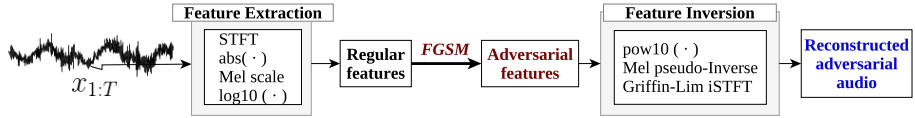}
  \caption{AAEs generation via feature inversion applied to the adversarial features}
  \label{fig:feat_inv}
\end{figure}

\paragraph{\textbf{Adversarial Audio Generation.}}
The four trained ASR networks are subsequently integrated with an algorithm called Fast Gradient Sign Method (FGSM ~\cite{Goodfellow2015}) adapted to the hybrid CTC-attention ASR system, with a focus on attention-decoded sequences~\cite{Chavez2020}.
It generates an adversarial instance for each utterance of the four \textit{non-adversarial} test sets with different degrees of MP3-compression.
This method is similar to the sequence-to-sequence FGSM, as proposed in~\cite{Sun2019}.
A previously decoded label sequence $y_{1:L}^*$ is used as reference to avoid label leaking~\cite{kurakin2016adversarial}.
Using back propagation, cross-entropy loss $ J(x_{1:T}, y^*_l; \theta)$ is obtained from a {whitebox} model (whose parameters $\theta$ are fully known).
Gradients from a sliding window with a fixed length $l_w$ and {stride} $ \nu $ are accumulated to $  \nabla_{\text{AAE}}(x_t) $ (Eq. \ref{eq:gradient_sum}). 
Gradient normalization~\cite{dong2018boosting} is applied for accumulation of gradient directions.
The intensity of the AN (denoted by $\delta_{AAE}$) is determined by the $\epsilon$ factor in Eq. \ref{eq:FGSM}, which we set to 0.3. 
$\delta_{AAE}$ is then added to the original \textit{feature-domain} input $x_t$ (Eq.~\ref{eq:deltamin}), in order to trigger the network $f$ to output a wrong transcription, different from the ground truth $y$ (Eq. \ref{eq:wrong_output}).
\begin{align}
    \nabla_{\text{AAE}}(x_t) &= \sum_{i = 0}^{\lceil L/\nu \rceil} \left( 
    \frac{\nabla_{x_t}  \sum\limits_{l = i\cdot\nu}^{l_w} J(x_{1:T}, y_l^*; \theta ) }
    {||\nabla_{x_t} \sum\limits_{l = i\cdot\nu}^{l_w} J(x_{1:T}, y_l^*; \theta )||_1}
    \right), \quad  l\in [0;L] \label{eq:gradient_sum}\\
    \delta_{AAE}(x_t) &= \epsilon\cdot \sgn( \nabla_{\text{AAE}}(x_t) )     \label{eq:FGSM}\\
    \hat{x}_t &= x_t + \delta_{AAE}(x_t), \quad  \forall t\in[1,T]. \label{eq:deltamin} \\
    y &\neq f(\hat{x_{t}},\theta) \label{eq:wrong_output} 
\end{align}
Notably, we did not aim for psychoacoustically-optimized AN, i.e., adversarial noise that would be totally inaudible,
because the focus was not on developing powerful adversarial attacks, but rather on exploring ways to improve ASR models' robustness to more simplistic AN.
Because the networks take acoustic feature vectors as input, FGSM originally creates AEs in the {feature domain}.
Yet, in order to evaluate our research hypothesis, we needed AEs in the {audio domain}, that is, AAEs.
Hence, we proceeded to {invert} the adversarial features and thus obtain synthetically \textit{reconstructed} adversarial audio (rAAEs).
The exact steps for both forward feature extraction, as well as feature inversion are illustrated in Fig.~\ref{fig:feat_inv} and were implemented with functions from Librosa toolbox~\cite{mcfee2015librosa}.

Additionally, mind that {log-mel fbank} features are a lossy representation of the original audio input, since they lack the phase information of the spectrum. This in turn hinders a highly accurate audio reconstruction.
In fact, mere listening revealed that it was relatively easy to distinguish the reconstruction artefacts when comparing pairs of original (non-adversarial) audio samples with their reconstructed versions\footnote{Reconstructed audio samples (both non-adversarial and adversarial) can be retrieved from~\url{https://github.com/iustinaabc/ASR-mp3-compression-AAEs}}.
Yet to see whether the reconstruction method impaired in any way the ASR performance, we performed the following sanity check:
the ASR model trained on uncompressed \textit{.wav} files was used to decode both the original test set and its reconstructed counterpart (obtained by feature extraction directly followed by inversion, i.e., without applying FGSM).
We observed just a mild 1.3\% absolute increase in the Character Error Rate (CER) for the reconstructed set, which implies that the relevant acoustic features are still accurately preserved following audio reconstruction with our feature inversion method.

\section{Results and Discussion}

\paragraph{\textbf{ASR results for non-adversarial vs. adversarial input.}}
Table \ref{tab:CER_adv_input} conveys the ASR results of decoding various test sets that originate from the same audio format as the training data of each ASR model, hence in a train-test \textit{matched} setting.
Adversarial inputs in the feature domain (column [b]) render far more transcription errors (by an absolute mean of $+52.05\%$ over all models) compared to the {baseline}, non-adversarial input listed in column [a].
This validates the FGSM method as effective in creating adversarial features from input data of any audio format (uncompressed, as well as MP3-compressed).
The error scores for reconstructed AAEs (rAAEs in column [c]) created from the adversarial features are also higher than the baseline, but, interestingly, lower than the CER scores for the adversarial features themselves.
This suggests that our reconstruction method makes the AAEs less powerful in misleading the model, which on the other hand is beneficial for the system's robustness to AN.
When we further compress the rAAEs with MP3 at the bitrate of $24$ kbps (column [d]), we observe an additional decline in the CER, thus indicating that MP3 compression is favourable in reducing the attack effectiveness of the adversarial input.
However, these CER values are still much higher than the baseline by a mean absolute difference of $+38.05\%$ across all ASR models, suggesting that the original transcription could \textit{not} be fully recovered.
The strongest reduction effect ($-21.31\%$) between MP3 rAAEs and the ``original'' adversarial features can be observed in the case of adversarial compressed input (24 kbps) originating from 64 kbps compressed data.
Overall, the mere numbers show that MP3 compression manages to {partially} reduce the error rates to adversarial inputs.
\begin{table}[!tb]
\centering
\caption{CER results for decoding adversarial input (marked as [b], [c] and [d]).
The last column indicates a relative CER score difference between adversarial features [b] and reconstructed, MP3-compressed AAEs [d], calculated as $\frac{[d]-[b]}{[b]} \cdot 100 (\%)$.
The inputs from column [d] were compressed at 24 kbps.}
\label{tab:CER_adv_input}
\resizebox{\textwidth}{!}{%
\begin{tabular}{@{}c|cccc|c@{}}
\toprule
\multirow{2}{*}{\textbf{\begin{tabular}[c]{@{}c@{}}ASR model \\ (source format of \\ train \& test inputs in \\ {[}a{]}, {[}b{]}, {[}c{]}, {[}d{]})\end{tabular}}} &
  \multicolumn{4}{c|}{\textbf{Input test data \& corresp. CER scores}} &
  \multirow{2}{*}{\textbf{\begin{tabular}[c]{@{}c@{}} \\ Relative CER \\ difference (\%) \\ between {[}b{]} and {[}d{]}\end{tabular}}} \\ \cmidrule(lr){2-5}
 &
  \textbf{\begin{tabular}[c]{@{}c@{}} {[}a{]} \\ Original \\ audio \end{tabular}} &
  \textbf{\begin{tabular}[c]{@{}c@{}} {[}b{]} \\ Adv. \\ features \end{tabular}} &
  \textbf{\begin{tabular}[c]{@{}c@{}} {[}c{]} \\ rAAEs \\ (reconstructed) \end{tabular}} &
  \textbf{\begin{tabular}[c]{@{}c@{}} {[}d{]} \\ MP3  \\rAAEs \end{tabular}} &
   \\ \midrule
uncompressed (\#1) & $17.8$ & $70.5$ & $62.2$ & $57.4$ & $-18.58$ \\
128 kbps-MP3 (\#2) & $18.8$ & $72.3$ & $64.0$ & $58.4$ & $-19.23$ \\
64 kbps-MP3 (\#3)  & $18.6$ & $71.8$ & $63.1$ & $56.5$ & $\mathbf{-21.31}$ \\
24 kbps-MP3 (\#4)  & $20.2$ & $69.0$ & $60.5$ & $55.3$ & $-19.86$ \\ \bottomrule
\end{tabular}}
\end{table}

\begin{figure}[!tb]
  \centering
\includegraphics*[width=0.82\linewidth]{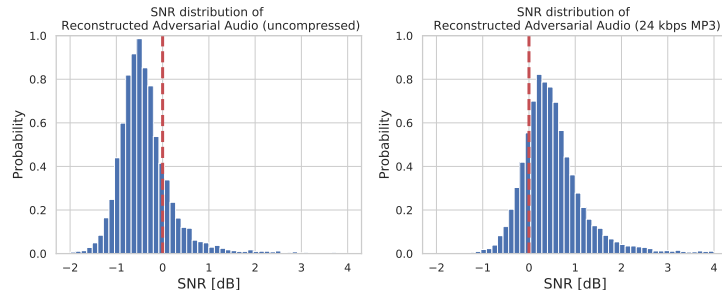}
   \caption{Normalized histograms following SNR estimation of reconstructed adversarial inputs (rAAEs): uncompressed (left) and MP3 at 24 kbps (right). The number of histogram bins was set to 50 for both plots.}
  \label{fig:SNRhistogram}
\end{figure}

\paragraph{\textbf{MP3 effects on Signal-to-Noise Ratio (SNR).}}
SNR is a measure that quantifies the amount of noise in an audio sample.
As such, it comes natural to assess how MP3 encoding impacts the AN in terms of SNR.
Since it was difficult to compute the SNR in the original, adversarial features' domain, we estimated the SNR of the reconstructed AAEs before and after MP3 compression as follows:
\begin{align}
    {\text{SNR}}_{\text{adv}}  &= 10 \log_{10} \frac{\text{power of the reconstructed speech (non-adversarial)}}{\text{power of the reconstructed AN}}
\end{align}
For both the uncompressed and MP3 rAAEs, the SNR was calculated with reference to the same signal in the numerator:
the reconstructed version of the \textit{original} speech utterance (no FGSM), so as to introduce similar artifacts as when reconstructing the adversarial audio, and thus have a more accurate SNR estimation.
As expected, we obtained different SNRs for each adversarial audio, because AN varies with the speech input.
Moreover, the SNR values differed before and after MP3 compression for each adversarial sample.
As the normalized histograms in Fig.~\ref{fig:SNRhistogram} show, most of the SNR values for the uncompressed rAAEs are negative (left plot), suggesting that the added AN has high levels and is therefore audible.
However, after MP3 compression (right plot), most adversarial samples acquire positive SNRs, implying that AN was diminished due to MP3 encoding.
To validate this, we applied the non-parametric, two-sided Kolmogorov-Smir\-nov (KS) statistical test~\cite{KS_stat_test_2} to evaluate whether the bin counts from the normalized SNR histograms of rAAEs before and after MP3 compression originate from the same underlying distribution.
We obtained a \emph{p-value} of 0.039 ($< 0.05$), 
confirming  the observed difference between the underlying distributions of the two histograms as statistically significant.
Thereby, MP3's incremental effect on the SNR values of adversarial samples was validated, which essentially means that MP3 reduces the AN.

\paragraph{\textbf{ASR results for input augmented with regular noise.}}
To have a reference for the behaviour of the ASR systems when exposed to common types of noise, we assessed the effects of MP3 compression on audio inputs corrupted by regular, non-adversarial noise as well.
For this complementary analysis, we augmented the original test sets with four noise types, namely \emph{white}, \emph{pink}, \emph{brown} and \emph{babble} noise (overlapping voices), each boasting distinct spectral characteristics.
The noise-augmented samples and their MP3-compressed versions were then fed as input to the decoding stage of the corresponding ASR system, i.e., the one that was trained on audio data of the same format as the test data.
Table~\ref{tab:noiseaugm} lists the CER results of this decoding experiment performed by ASR model \#1 (trained on uncompressed data).
\begin{table}[!tb]
\centering
\caption{CERs for test sets augmented with regular noise (in uncompressed and MP3 formats) of different SNR values, decoded by ASR model \#1}
\label{tab:noiseaugm}
\setlength{\tabcolsep}{4pt}
\begin{tabular}{@{}c|c c c c c c@{}}
\toprule
\multirow{2}{*}{\textbf{Test set augmented with:}} & \multicolumn{6}{c}{\textbf{SNR {[}dB{]}}}        \\ \cmidrule(l){2-7} 
   & \textbf{30}  & \textbf{10}   & \textbf{5}    & \textbf{0}    & \textbf{-5}   & \textbf{-10}  \\ \midrule
\textbf{{[}A{]} white noise} & 19.1 & 32.7 & 41.9 & 53.7 & 66.2 & 78.2 \\
\textbf{ MP3 compressed (24kbps)}  & 29.1 & 51.2 & 61.7 & 71.2 & 78.7 & 86   \\ \midrule
\textbf{{[}B{]} pink noise} & 18.5 & 29.1 & 38.1 & 51.7 & 67.4 & 82.1 \\
\textbf{ MP3 compr. (24kbps)} & 26.9 & 42.5 & 53   & 66.4 & 79.8 & 89.9 \\ \midrule
\textbf{{[}C{]} brown noise} & 17.9 & 19.7 & 21.9 & 26.1 & 34.1 & 47.8 \\
\textbf{ MP3 compr. (24kbps)} & 25.3 & 29   & 32.5 & 38   & 47.3 & 60.6 \\ \midrule
\textbf{{[}D{]} babble noise} & 18.3 & 35.8 & 53.4 & 77.4 & 89   & 93.6 \\
\textbf{ MP3 compr. (24kbps)} & 25.8 & 48.2 & 66   & 83.6 & 93.1 & 95.4 \\ \bottomrule
\end{tabular}
\end{table}
Mind that all ASR systems were trained on the original, noise-free data; therefore,
decoding noisy inputs was expected to cause more transcription errors than the original, clean data.
Based on CER, one can observe that the lower the SNR (or the higher level of noise added to the input), the more error-prone the ASR system is, irrespective of the noise type.
The most adverse effect seems to be in the case of white and babble noise, rows [A] and [D], which also happen to have the richest spectral content, imminently interfering with the original speech.

Yet of utmost interest is what happens when the {same treatment} used for adversarial inputs, i.e., MP3 compression, is applied to the novel speech inputs enhanced with regular noise.
These results are illustrated every second row in Table~\ref{tab:noiseaugm}.
Error rates turn out always higher for MP3-compressed inputs than for the uncompressed ones, regardless of noise type or SNR.
Consequently, MP3 compression has the \emph{inverse effect} compared to what was observed for adversarial noise:
while it triggered a {reduction} in the amount of errors returned by the ASR systems to {adversarial input}, it {failed} to do so for {non-adversarial noise}.
On the contrary, MP3 compression {increased} the amount of errors for inputs augmented with non-adversarial noise, especially at high and mid-range SNRs.
This further validates that MP3 compression has the desired effect of {partially} reducing only adversarial perturbations, whereas deteriorating the non-adversarial, regular noise.

\section{Conclusion}
In this work, we explored the potential of MP3 compression as a countermeasure
against Adversarial Noise compromising speech inputs.
To this end, we constructed adversarial feature samples with the FGSM gradient-based method. 
The adversarial features were then mapped back into the audio domain by inverse feature extraction operations. 
The resulting adversarial audio (denoted as \emph{reconstructed}) was thereafter MP3-compressed and presented to the input of four ASR models featuring a hybrid CTC-attention architecture,
having been previously trained on four types of audio data.
\textbf{Our three key findings are}:
\begin{enumerate}
    \item
    In comparison to \emph{adversarial features}, reconstructed AAEs, as well as MP3 compressed AAEs had lower error rates in their transcriptions.
    The error reduction did not achieve the performance for non-adversarial input, which implies that correct transcriptions were not completely recovered. 
    \item 
    MP3 compression on the estimated SNR values of rAAEs yielded a statistically significant effect, supporting the observation that MP3 increased the SNR values of adversarial samples, which translates to AN reduction.
    \item
    Our experiments with \emph{non-adversarial} noise suggest that MP3 compression is beneficial only in mitigating Adversarial Noise, while it deteriorates the speech recognition performance to non-adversarial noise.
\end{enumerate}

%
%
\bibliographystyle{splncs04}
\bibliography{paper}

\end{document}